\documentclass[aps,pra,twocolumn,superscriptaddress,amsmath,amssymb,preprintnumbers,showpacs]{revtex4}

\usepackage{graphicx}
\usepackage{dcolumn}

\begin{document}


\title{Non-Markovian dynamics of a qubit}

\author{Sabrina Maniscalco}
\email{maniscalco@ukzn.ac.za}

\affiliation{School of Physics, University of KwaZulu-Natal,
Durban 4041, South Africa}


\author{Francesco Petruccione}
\email{petruccione@ukzn.ac.za}

\affiliation{School of Physics, University of KwaZulu-Natal,
Durban 4041, South Africa}

\begin{abstract}
In this paper we investigate the non-Markovian dynamics of a qubit
by comparing two generalized master equations with memory. In the
case of a thermal bath, we derive the solution of the recently
proposed post-Markovian master equation [A. Shabani and D.A.
Lidar, Phys. Rev. A {\bf 71}, 020101(R) (2005)] and we study the
dynamics for an exponentially decaying memory kernel. We compare
the solution of the post-Markovian master equation with the
solution of the typical memory kernel master equation. Our results
lead to a new physical interpretation of the reservoir correlation
function and bring to light the limits of usability of master
equations with memory for the system under consideration.

\end{abstract}

\pacs{03.65.Yz,03.65.Ta,42.50.Lc}

\maketitle

\section{Introduction}
The dynamics of systems interacting with their surroundings is in
general very complicated. Very often, however, the physical
systems of interest are sufficiently isolated from their
environment to allow the use of certain approximations such as the
weak coupling approximation and the Markovian approximation
\cite{petruccionebook}. The former one assumes that the
interaction between the system and the environment is sufficiently
weak, i.e. the system is quasi-closed. The latter one relies on
the assumption that the characteristic times of the system are
much longer than those of the environment, and it always assumes
the validity of the weak coupling approximation.

Most of the results on open systems dynamics are based on the weak
coupling and Markovian approximations. Recent studies have shown
the limits of the Markovian description of quantum computation and
quantum error correction
\cite{Alicki02,Ahn,Daffer04,Terhal05,Preskill}. Moreover,
nanotechnology-based devices using hybrid systems, e.g. combining
quantum optical and solid state systems, have been investigated
and seem to be very promising for future technological
applications \cite{Tian04,Tian04b}. In order to describe
decoherence in many solid state systems non-Markovian approaches
need often to be used \cite{John94,quang97,Vega05,Florescu04}.
Finally,  a comprehensive and complete understanding of the
interaction between a quantum system and its environment, not
relying on the weak coupling and/or Markovian approximations, is
crucial in order to clarify fundamental issues such as the
quantum-classical border, and in order to gain new insight in the
dynamics of quantum systems which are not in thermal equilibrium.

Outside the region of validity of the Markovian approximation the
master equation describing the dynamics cannot be usually cast in
the well known Lindblad form \cite{Lindblad,Gorini}. This fact has
several consequences: one for all, complete positivity of the
dynamical map \cite{note1} is not guaranteed anymore, and even
positivity may be violated. The latter of this properties, i.e.
positivity, is necessary to guarantee the statistical
interpretation of the density matrix, while the former one ensures
that the time evolution in the system-environment total space is
unitary .

Non-Lindblad master equations are much more difficult to solve,
both analytically and numerically, than Lindblad ones, and they
may lead to non-physical behaviors such as, e.g., violation of
positivity of the dynamical map [See, e.g., Refs.
\cite{Munro96,barnett,Maniscalco05}]. The break down of the
positivity condition stems from the phenomenological nature of
most of the non-Markovian approaches. Exact generalized master
equations indeed, by definition, do not violate neither positivity
nor complete positivity but they are generally far too complicated
for providing useful means for studying the system dynamics.

In this paper we focus on a basic model of an open quantum system
with memory, i.e. a two-level system (qubit) interacting with a
bosonic thermal reservoir. We apply a recently proposed
post-Markovian approach \cite{lidar05} which interpolates between
the exact Kraus map and the Markovian dynamics. We compare the
solution of the post-Markovian generalized master equation with
the solution of the typical memory kernel master equation
\cite{petruccionebook,barnettbook}. For a specific physically
interesting form of the reservoir spectral density it is possible
to derive an exact solution starting from a microscopic
description of the total system, i.e. system plus reservoir. This
fact allows us to make a comparison between the phenomenological
memory kernel approaches and the exact microscopic approach. From
this comparison a new physical interpretation of the reservoir
correlation function will emerge. Finally the limitations of the
memory kernel approaches will be underlined and issues related to
the loss of positivity of the dynamical maps will be carefully
analyzed.

The paper is structured as follows. In Sec. II we recall the
post-Markovian master equation and we present the solution for the
case of a qubit interacting with a thermal reservoir. In Sec. III
we recall the solution of the typically used generalized master
equation with memory, we compare it to the post-Markovian
solution,  and we analyze the \lq\lq non-physical\rq\rq region of
the parameters space where positivity breaks down. In Sec. IV we
consider the exact solution and in Sec. V we present conclusions.

\section{Post-Markovian master equation for a qubit}
\subsection{Interpolating the Kraus and Markovian dynamical maps}
Very recently a new general post-Markovian master equation has
been presented \cite{lidar05}. An interesting feature of this
phenomenological master equation is that, by construction, it
interpolates between the generalized measurement interpretation of
the exact Kraus operator sum map, and the continuous measurement
interpretation of the Markovian dynamics.

It is worth reminding that the dynamics of open quantum systems
may be described equivalently either by means of the density
matrix satisfying the master equation or by means of a stochastic
wave function which is the solution of the stochastic
Schr\"{o}dinger equation unravelling the dynamics
\cite{petruccionebook}. For Markovian dynamics there exist a
physical interpretation for the time evolution of the stochastic
wave functions (quantum trajectories). Indeed it has been shown
that the quantum trajectories describe the time evolution of the
system conditioned to a continuous measurement of the environment
\cite{carmichaelbook,Moelmer96}. Different types of detection
schemes of the environment (photon counting, homodyne and
heterodyne detection) correspond to different unravellings
(different types of stochastic Schr\"{o}dinger equations). At
contrast, in the case of non-Markovian dynamics, it has been shown
that quantum trajectories do not have a physical interpretation
\cite{Gambetta02} although attempts to find an interpretation in
extended Hilbert states have been performed \cite{Breuer04}. In
more detail, it turns out that the stochastic wave function at
time $t$ represents the state the system would be in at that time
if a measurement was performed on the environment at that time,
and yielded a particular result (generalized measurement
interpretation). However, the wavefunction at time $t$ does not
have any link with itself at times less then $t$, and therefore
there cannot be any physical interpretation of the quantum
trajectories, for non-Markovian systems \cite{Gambetta02}. The
reason can be traced back to the finite correlation time
characterizing the non-Markovian bath.

Now, since the post-Markovian master equation actually
interpolates between the exact dynamics and the Markovian
dynamics, an analysis of the time evolution according to this
equation may give new insight in the dynamics of non-Markovian
systems, in the possible physical phenomena taking place when the
Markovian approximation fails, and in their role in the break down
of the continuous measurement interpretation. This is in fact what
we are going to investigate in the rest of the paper. Moreover, we
will study the usefulness of the post-Markovian master equation
presented in Ref. \cite{lidar05} for the description of open
quantum systems, comparing it to other common non-Markovian
approaches. To this aim we consider the basic open quantum system,
e.g. a two-level atom or qubit, interacting linearly with a
quantized bosonic reservoir at $T$ temperature.

\subsection{Post-Markovian master equation for the qubit}
The general form of the post-Markovian master equation introduced
in Ref. \cite{lidar05} is the following
\begin{equation}
\frac{d \rho}{d t} = \mathcal{L} \int_0^t dt' k(t')
\exp(\mathcal{L}t')\rho(t-t'), \label{eq:1}
\end{equation}
where $\rho(t)$ is the density matrix of the reduced system,
$k(t')$ is the memory kernel, and $\mathcal{L}$ is the Markovian
Liouvillian. For the system here considered the Markovian
Liouvillian is given by \cite{GardinerZollerbook}
\begin{eqnarray}
\mathcal{L} \rho &=& \gamma_0 (N+1) \left[ \sigma_- \rho \sigma_+
- \frac{1}{2} \sigma_+ \sigma_- \rho - \frac{1}{2}
\rho \sigma_+ \sigma_- \right] \nonumber\\
&+& \gamma_0 N \left[ \sigma_+ \rho \sigma_- - \frac{1}{2}
\sigma_- \sigma_+ \rho - \frac{1}{2} \sigma_- \sigma_+
\right],\label{eq:markovme}
\end{eqnarray}
with $\gamma_0$ the phenomenological dissipation constant, $N$ the
mean number of excitations of the reservoir, and $\sigma_{\pm}$
the spin inversion operators. In Appendix A we recall the form of
the solution of the Markovian master equation via the damping
basis method.

In the following we will focus on a widely use form of memory
kernel, namely the exponential memory kernel
\begin{equation}
k(t)=\gamma e^{- \gamma t}. \label{eq:kernel}
\end{equation}
It is worth stressing that $k(t)$, which hereafter we will call
the Shabani-Lidar memory kernel, is a quantity introduced
phenomenologically in Ref. \cite{lidar05}. Therefore, it should
not be confused with the memory function, appearing in the second
order approximation of the Nakajima-Zwanzig equation, which is
related to the spectral density of the reservoir [See, e.g.,
\cite{petruccionebook}, p.465].  In order to understand the
meaning of the Shabani-Lidar memory kernel we recall that, in the
measurement scheme approach to open quantum systems, the
post-Markovian master equation describes a situation in which a
measurement of the environment at a time $t'$ is followed by a
Markovian evolution, described in terms of continuous measurements
of the environment at times $t>t'$. The time $t'$ characterizes
the bath memory effects. In this picture, the Shabani-Lidar memory
kernel is a function which assigns weights to different
measurements (selecting different $\rho(t')$) \cite{lidar05}. Now,
having in mind these definitions, and remembering that that the
post-Markovian master equation is phenomenological, one might
wonder which is the relationship between the memory kernel of the
post-Markovian master equation and the Nakajima-Zwanzig memory
function (also known as correlation function) which is related to
the reservoir spectral density. This question will be addressed in
Sec. IV.

\subsection{Analytic solution}
By applying the method described in \cite{lidar05}, and recalled
in Appendix B, to a two-level system whose ground and excited
states are $\vert 1 \rangle $ and $\vert 2 \rangle$, respectively,
we derive the following solution of the post-Markovian master
equation
\begin{eqnarray}
\rho(t) &=& \frac{1}{2} \left\{ I - \left[ \xi(R,t) \left(
\rho_{11}-\rho_{22}+\frac{1}{2N+1}\right) \right. \right.\nonumber \\
&-& \left. \left.  \frac{1}{2N+1}\right] \sigma_z +
\frac{\xi(2R,t)}{2} \left[ \rho_{12} \sigma_+ + \rho_{21} \sigma_-
\right] \right\}, \label{eq:solutionpostM}
\end{eqnarray}
where $\rho_{ij} = \langle i \vert \rho(0) \vert j \rangle$, with
$i,j=1,2$. In the previous equation
\begin{eqnarray}
\xi(R,\tau)&=& e^{-\frac{R+1}{2} \tau} \left\{
\frac{1}{\sqrt{|1-r(R)|}} \sinh \left[\sqrt{|1-r(R)|}\frac{(R+1)\tau}{2}\right] \right. \nonumber \\
&+& \left. \cosh \left[ \sqrt{|1-r(R)|}\frac{(R+1)\tau}{2}\right]
\right\}, \label{eq:xi2}
\end{eqnarray}
and
\begin{equation}
r(R) = \frac{4 R}{(R+1)^2}, \label{eq:r2}
\end{equation}
with $R= |\lambda_2|/\gamma$, the eigenvalue $\lambda_2$ being the
one derived for the Markovian master equation [see Eq.
(\ref{eq:lambdai})], and $\tau = \gamma t$.

Equation(\ref{eq:xi2}) is valid only for $r(R) \le 1$ and $r(2R)
\le 1$. When $r(R) > 1$ and/or $r(2R) >1$, the form of the time
dependent coefficients $\xi(R,\tau)$ and/or $\xi(2R,\tau)$
appearing in Eq. (\ref{eq:solutionpostM}) is obtained from Eq.
(\ref{eq:xi2})by substituting $\sinh[.]$ and $\cosh[.]$ with
$\sin[.]$ and $\cos[.]$, respectively.

Let us focus on the case of a zero-temperature reservoir. In this
case the solution of the post-Markovian master equation takes the
form
\begin{eqnarray}
\rho(\tau)= \frac{1}{2} \{ I+  \left[
 2 P_1(\tau) -1\right]
 \sigma_z + \left[ 2
\rho_{12}(\tau) \sigma_+ + h.a. \right] \},
\end{eqnarray}
where the time evolution of the ground state population
$P_1(\tau)= \rho_{11}(\tau)$  and of the coherences
$\rho_{12}(\tau) = \rho_{21}^*(\tau)$ is given, respectively, by:
\begin{eqnarray}
P_1(\tau)&=& \rho_{11} \xi(R,\tau), \label{eq:p1postMa} \\
\rho_{12}(\tau) &=& \frac{1}{4}\rho_{12} \xi(2R,\tau).
\end{eqnarray}
We note that, for the zero-temperature case here considered,
$\lambda_2 = - \gamma_0$, and therefore $R=\gamma_0/\gamma$.

\subsection{Markovian limit}
We conclude this section by showing the Markovian limit of the
post-Markovian solution. To this aim we firstly notice that Eq.
(\ref{eq:xi2}) can be recast in the following simplified form
\begin{eqnarray}
\xi(R,t) &=& \frac{e^{- |\lambda_2|t} - R e^{- \gamma t}}{1-R}.
\label{eq:xipostM}
\end{eqnarray}
For times $t \gg 1/\gamma \equiv \tau_R$ (coarse graining in time)
and for $R \ll 1$, i.e. $\tau_0 = 1/\gamma_0 \gg \tau_R$, the
previous equations become $\xi(R,t) \simeq e^{- |\lambda_2|t}$,
while $\xi(2R,t) \simeq e^{- 2|\lambda_2|t}$, and one reobtains
the Markovian dynamics. The approximation $\tau_0 = 1/\gamma_0 \gg
\tau_R$ amounts at saying that $\tau_R$ is much smaller than the
characteristic time of the system $\tau_0$. In the following we
will call $\tau_R$ the characteristic time of the reservoir.

\section{Generalized master equation with memory}
\subsection{Phenomenological master equation}\label{subsec:phenomenologic}
Let us now consider the typical phenomenological  master equation
with memory kernel having the form
\begin{equation}
\frac{d \rho}{dt} = \int_0^t dt' k(t') \mathcal{L}  \rho(t-t').
\label{eq:memoryME}
\end{equation}
This generalized master equation takes into account the previous
\lq\lq history\rq\rq  ($0<t'<t$) of the density matrix $\rho(t)$
by means of the phenomenological memory kernel $k(t')$.
Differently from time-convolutionless approaches
\cite{petruccionebook}, this leads to a master equation which is
not local in time. For specific forms of the memory kernel, as the
exponential one considered in this paper, this master equation can
be solved by means of the Laplace transforms.

The memory kernel master equation given by Eq. (\ref{eq:memoryME})
has been used in Ref. \cite{barnett} for studying the
non-Markovian dynamics of a quantum harmonic oscillator
interacting with the vacuum. There the authors have found that,
for an exponential memory kernel such as the one given by Eq.
(\ref{eq:kernel}), the positivity of the density matrix is
violated for certain values of the phenomenological decay
constants. In Ref. \cite{Daffer04} the generalized master equation
with memory given in  Eq. (\ref{eq:memoryME}) has been analyzed
for a two-level atom in the presence of telegraphic noise, and
conditions for complete positivity have been presented. In the
present paper we consider this model for the two-level atom
interacting with a bosonic reservoir at $T$ temperature, focussing
in particular, for the sake of simplicity, on the zero temperature
reservoir.

Similarly to the case of the post-Markovian master equation, one
can solve Eq. (\ref{eq:memoryME}) by taking its Laplace transform,
determining the poles, and inverting the solution in the standard
way. Using the damping basis given by Eq. (\ref{eq:dampingbasis}),
with the eigenvalues given by Eq. (\ref{eq:lambdai}), we find that
the solution has the same form of Eq. (\ref{eq:solutionpostM}),
with  the only difference that the quantities $\xi(R,t)$ is now
given by
\begin{eqnarray}
\xi(R, \tau)&=& e^{-\frac{\tau}{2}} \left\{
\frac{1}{\sqrt{|1-4R|}} \sinh \left[ \frac{\tau}{2} \sqrt{|1-4R|}
\right] \right. \nonumber \\ &+& \left. \cosh \left[
\frac{\tau}{2} \sqrt{|1-4R|}
 \right] \right\}. \label{eq:xi2memory}
\end{eqnarray}
By comparing Eq. (\ref{eq:xi2}) and Eq. (\ref{eq:xi2memory}), with
the help of Eq. (\ref{eq:r2}), one easily sees that this amounts
at assuming $R= \gamma_0/\gamma \ll 1$ in Eq. (\ref{eq:xi2}).
Stated another way, as one can see directly form Eqs. (\ref{eq:1})
and (\ref{eq:memoryME}), the memory kernel master equation is a
special case of the post-Markovian master equation, in the limit
in which the system characteristic time $\tau_0$ is much bigger
than the reservoir correlation time $\tau_R$ \cite{lidar05}. We
note that, for $4R>1$ and $8R > 1$ the form of the time dependent
coefficients $\xi(R,\tau)$ and/or $\xi(2R,\tau)$, respectively, is
obtained from Eq. (\ref{eq:xi2memory}) by substituting $\sinh[.]$
and $\cosh[.]$ with $\sin[.]$ and $\cos[.]$.

We conclude this section noting that, as for the Markovian
dynamics, the solution of the master equation with memory, given
by Eq. (\ref{eq:memoryME}), can be obtained from the solution of
the post Markovian master equation in the limit $R=
\gamma_0/\gamma \ll 1$. However, contrarily to the Markovian
dynamics, no coarse graining in time has been made, and therefore
the solution of Eq. (\ref{eq:memoryME}) describes correctly the
short time dynamics, $t \ll \tau_R$, characterized by
non-negligible system-reservoir correlations.

\subsection{Positivity of the dynamical map}\label{sec:positivity}
As we have already mentioned in the Introduction, since the master
equations given by Eqs. (\ref{eq:1}) and (\ref{eq:memoryME}) are
not of Lindblad-type, the positivity of their corresponding
dynamical maps is not guaranteed. The break down of the positivity
condition means that the density matrix loses its statistical
interpretation, its eigenvalues becoming negative. This is hence a
sign of failure of the phenomenological master equations with
memory.

An analysis of the positivity condition for a master equation of
the form of Eq.(\ref{eq:memoryME}) has been presented, in the case
of a damped harmonic oscillator, in Ref. \cite{barnett}. There the
authors have found that positivity is always violated for
sufficiently high values of the phenomenological decay constant.
The positivity condition of the memory kernel master equation has
been also studied in Ref. \cite{Budini04} for different types of
memory kernels, including the exponential one, generalizing the
results of \cite{barnett}. It is therefore not surprising that, as
we will see in the following, we obtain the same result of Ref.
\cite{barnett} for the same form of master equation, with an
exponential memory kernel, when the system is a qubit. As we will
show, however, the post-Markovian master equation exhibits a
strikingly different behavior in respect of the positivity
condition.

In order to study in more detail the positivity condition for both
the two memory kernel master equations considered in this paper,
it is more convenient to rewrite the solutions in terms of the
Bloch vector $\vec{w}= \{ w_x,w_y,w_z \}$. The qubit density
operator at time $\tau = \gamma t$, given by
Eqs.(\ref{eq:solutionpostM}) and Eq. (\ref{eq:xi2}) in the
post-Markovian case, and by Eqs.(\ref{eq:solutionpostM}) and
(\ref{eq:xi2memory}) in the memory kernel case, can be recast in
the form
\begin{equation}
\rho(\tau)= \frac{1}{2} \left[ I+ \vec{w}(\tau) \cdot \vec{\sigma}
\right],
\end{equation}
with $\vec{\sigma} = \{ \sigma_x, \sigma_y, \sigma_z\}$, and
\begin{eqnarray}
w_x(\tau) &=& \xi (2R, \tau) {\rm Re}[\rho_{12}] = \xi (2R, \tau) w_x(0) \label{eq:wx} \\
w_y(\tau) &=& - \xi (2R, \tau) {\rm Im}[\rho_{12}] = \xi (2R, \tau) w_y(0) \label{eq:wy}\\
w_z(\tau) &=& 2P_1 \xi(R,\tau) -1 = P_1(\tau)-P_2(\tau),
\label{eq:wz}
\end{eqnarray}
where $P_1(\tau)$ is given by Eq.(\ref{eq:p1postMa}),
$P_2(\tau)=1-P_1(\tau)$, and $\xi(R,\tau)$ is given by Eq.
(\ref{eq:xi2memory}) (post-Markovian) or Eq. (\ref{eq:xi2})
(memory kernel). The dynamical map describing the evolution of the
qubit is positive if and only if the density operator evolves only
to states inside or on the Bloch sphere. Therefore the positivity
condition, in terms of the Bloch vector components, simply reads
$|w_i(\tau)|\le 1$, $i=x,y,z$. By looking at Eqs.
(\ref{eq:wx})-(\ref{eq:wz}) and Eq.(\ref{eq:p1postMa}), one sees
immediately that the conditions $|w_i(\tau)|\le 1$ are equivalent
to $|\xi(R,\tau)|\le 1$ and $|\xi(2R,\tau)| \le 1$.

Let us begin considering the post-Markovian master equation. It is
straightforward to prove that $\xi(R,\tau)$ is a positive and
monotonically decreasing function of $R$, therefore, it is
sufficient to investigate the conditions for which $|\xi(R,\tau)|
\le 1$, since $|\xi(2R,\tau)| \le |\xi(R,\tau)|$ for all values of
$R$ and $\tau$. As shown in Appendix C, it turns out that the
post-Markovian dynamical map never violates positivity. On the
contrary, for the case of the memory kernel master equation given
by Eq. (\ref{eq:memoryME}), a close look to Eq.
(\ref{eq:xi2memory}) tells us that $|\xi(R,\tau)| \le 1$ only for
$4R<1$, i.e. $4 \gamma_0 < \gamma$. We remind that one may derive
the memory kernel master equation given by Eq. (\ref{eq:memoryME})
from the post-Markovian master equation given by Eq. (\ref{eq:1})
in the limit $\gamma_0 \ll \gamma$. Having this is mind it is not
surprising that the positivity condition breaks down for
sufficiently high values of $\gamma_0$, for the system here
considered. The study of positivity, therefore, suggests that the
post-Markovian master equation is somehow \lq\lq more
fundamental\rq\rq than the memory kernel master equation.

\section{Exact dynamics}
In this section we will analyze two different aspects
characterizing the non-Markovian dynamics of a quibit described by
means of memory kernel master equations. The first aspect stems
from the microscopic derivation, using the Nakajima-Zwanzig
formalism, of the phenomenological memory kernel master equation
given by Eq. (\ref{eq:memoryME}). The microscopic derivation will
allow us to link the Shabani-Lidar memory kernel to the
correlation function and to the reservoir spectral density, in the
limit $\gamma_0 \ll \gamma$.

The second aspect described in this section is related to a
physically interesting specific microscopic  model of
non-Markovian dynamics of a qubit for which an exact solution does
exist. The existence of an exact solution allows to make a
comparison with the predictions of the post-Markovian and memory
kernel solutions, and hence to study the limits of both these
approaches. Moreover, keeping in mind that the post-Markovian
approach can be seen as an interpolation between the exact Kraus
map and the Markovian one, one may gain new insight in the reason
why non-Markovian quantum trajectories lack of a physical meaning.

\subsection{Microscopic derivation}
Let us begin with the microscopic derivation of the memory kernel
master equation given by Eq. (\ref{eq:memoryME}) . We consider the
following microscopic Hamiltonian for the total system, i.e.
system plus reservoir,
\begin{eqnarray}
H= H_0+H_I,
\end{eqnarray}
with
\begin{eqnarray}
H_0 &=& \frac{\hbar \omega_0}{2}\sigma_z + \sum_k \omega_k
b^{\dag}_k
b_k, \nonumber \\
H_I &=& \sigma_+ B + \sigma_- B^{\dag},\label{eq:HI}
\end{eqnarray}
where $B= \sum_k g_k b_k$. The transition frequency of the
two-level system is denoted by $\omega_0$, the index $k$ labels
the different modes of the reservoir with frequencies $\omega_k$,
$b^{\dag}_k$ and $b_k$ indicate the creation and annihilation
operators, and $g_k$ are the coupling constants. Following the
typical approach for the derivation of master equations for the
reduced density matrix we firstly write the von Neumann master
equation for the total system in the interaction picture, then we
trace over the environmental degrees of freedom, under the
assumptions that ${\rm Tr }_R \left[ H_I(t),\rho_T(0) \right]=0$,
with $\rho_T$ the density matrix of the total system, and
$\rho_T(0)=\rho(0)\otimes \rho_R(0)$, with $\rho_R(0)$ the density
matrix of the reservoir. To the second order in perturbation
theory (Born approximation), we obtain the following
integro-differential equation for the reduced density matrix
\begin{eqnarray}
\frac{d \rho(t)}{dt}=\int_0^t dt' {\rm Tr}_R
\left[H_I(t),[H_I(t'), \rho(t') \otimes \rho_R ]\right].
\label{eq:mastersecord}
\end{eqnarray}
Having in mind Eq.(\ref{eq:HI}) it is then straightforward to
recast Eq.(\ref{eq:mastersecord}) in the same form of Eq.
(\ref{eq:memoryME}) with
\begin{eqnarray}
k(t') &=& {\rm Tr}_R \left[ B(0)B^{\dag}(t') \rho_R \right]
\nonumber \\
&\equiv& \int d \omega J(\omega) e^{i (\omega-\omega_0)t'}.
\label{eq:kernel2}
\end{eqnarray}
In the previous equation $B(t) = \sum_k g_k b_k \exp (-i \omega_k
t )$ is the reservoir operator appearing in Eq. (\ref{eq:HI}), in
the interaction picture, and $J(\omega) = \sum_k g _k^2
\delta(\omega-\omega_k)/(2 m_k \omega_k)$, with $m_k$ masses of
the oscillators of the reservoir. From the previous definition of
$k(t')$ one sees clearly that this function describes temporal
correlations of the reservoir operators $B$, and therefore it is
commonly known as the reservoir correlation function. In the
second line of Eq. (\ref{eq:kernel2}) we have introduced the
so-called spectral density of the reservoir $J(\omega)$, which is
therefore simply the Fourier transform of the correlation
function. An exponential correlation function of the form of Eq.
(\ref{eq:kernel}) corresponds to a Lorentzian spectral density as
the one typical of cavity quantum electrodynamics, for a reservoir
in the vacuum state.

In the rest of this section we will focus on the following
specific physical system: a two level atom interacting resonantly
with a quantized mode of an empty high Q cavity. Assuming that the
two-level atom is in resonance with the cavity mode, the reservoir
spectral density is the following
\begin{equation}
J(\omega)= \frac{1}{\pi}
\frac{\bar{\gamma_0}\bar{\lambda}^2}{(\omega_0-\omega)^2+\bar{\lambda}^2}.
\label{eq:spectraldensity}
\end{equation}
Using Eq. (\ref{eq:kernel2}), we get
\begin{equation}
k(t')= \bar{\gamma}_0 \bar{\lambda} e^{- \bar{\lambda}t},
\label{eq:kernel3}
\end{equation}
and the master equation (\ref{eq:mastersecord}) becomes
\begin{equation}
\frac{d \rho}{dt} = \int_0^t dt' \bar{\mathcal{L}} k(t')
\rho(t-t'),
\end{equation}
with $k(t')$ given by Eq. (\ref{eq:kernel3}), and $\bar{\gamma}_0
\bar{\mathcal{L}}= (\bar{\gamma}_0/\gamma_0) \mathcal{L}$ , where
$\mathcal{L}$ is given by Eq. (\ref{eq:markovme}), with $N=0$. A
direct comparison with Eqs. (\ref{eq:kernel}) and
(\ref{eq:memoryME}) clearly shows that the master equation derived
using second order perturbation theory starting from the
microscopic description above coincides with the phenomenological
memory kernel master equation given by Eq.(\ref{eq:memoryME}),
provided that $\bar{\lambda}=\gamma$, $\bar{\gamma}_0 = \gamma_0$.

\subsection{Physical interpretation of the system and reservoir parameters}
The microscopic derivation allows to give a physical
interpretation of the decay constants $\gamma_0$ and $\gamma$.
Indeed, Eq. (\ref{eq:spectraldensity}) tells us that $\gamma_0$
measures the strength of coupling between the two-level atom and
the vacuum reservoir, and hence the system characteristic time
$\tau_0$ is determined only by the system-reservoir coupling
strength. The reservoir correlation time $\tau_R=1/\gamma$ is
simply given by the inverse of the width of the Lorentzian
spectral density.

Having this in mind, the reason of the violation of positivity for
the master equation given by Eq. (\ref{eq:memoryME}) becomes
clear. Indeed, such master equation correctly describes the
dynamics of the system only when second order perturbation theory
is valid, i.e. for weak system-reservoir coupling. The conditions
in correspondence of which positivity is violated, i.e. when
$4R>1$ and $8R>1$, with $R=\gamma_0/\gamma$, however, correspond
to strong couplings between system and reservoir.

Finally, we remind that, as noticed in Sec.
\ref{subsec:phenomenologic}, Eq.(\ref{eq:mastersecord}) is the
limit for $\gamma_0/\gamma \ll 1$ of the post-Markovian master
equation given by Eq.(\ref{eq:1}). As a consequence, in this
limit, the Shabani-Lidar phenomenological memory kernel $k(t')$
coincides with the Fourier transform of  the reservoir spectral
density, as given by Eq. (\ref{eq:kernel2}). This fact allows to
give a new physical interpretation to the correlation function of
the reservoir in terms of generalized measurement by the
environment. Indeed, it turns out that the reservoir correlation
function acts as a weighting time distribution function, assigning
weights to different measurements selecting different $\rho(t')$.

\subsection{Exact solution}

\begin{figure}
\includegraphics[width=8 cm,height=6 cm]{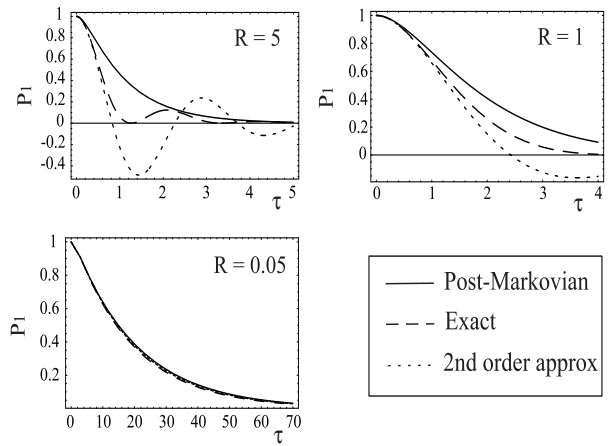}
\caption{Time evolution of the excited state probability for three
different values of $R$, i.e. $R=5$, $R=1$, $R=0.05$. The
dimensionless time is $\tau=\gamma t$. The solid line indicates
the dynamics for the post-Markovian master equation, the dashed
lines indicates the exact dynamics and the dotted line the
dynamics of the memory kernel master equation derived using second
order perturbation theory.} \label{fig:chi}
\end{figure}

The physical system we consider in this section is one of the few
open quantum systems amenable of an exact solution
\cite{Garraway97}. For a spectral density of the type given by Eq.
(\ref{eq:spectraldensity}), the exact density matrix has the
following form \cite{petruccionebook}
\begin{equation}
\rho(t) =
\begin{pmatrix}
  P_1(\tau) & \rho_{12}(\tau) \\
  \rho_{12}^*(\tau) & 1-P_1(\tau)
\end{pmatrix},
\end{equation}
with
\begin{eqnarray}
P_1(\tau) &=& P_1(0) e^{- \tau} \Big\{ \cosh \left[ \sqrt{|1-2R|}
\; \frac{\tau}{2} \right]  \nonumber
\\ &+&  \frac{1}{\sqrt{|1-2R|}} \sinh \left[ \sqrt{|1-2R|} \;
\frac{\tau}{2} \right] \Big\}^2, \label{eq:p1exact}
\end{eqnarray}
for $2R \le 1$. If $2R >1$, $P_1(t)$ has the same form of Eq.
(\ref{eq:p1exact}) provided that the substitutions $\cosh[.]
\rightarrow \cos[.]$ and $\sinh [.] \rightarrow \sin[.]$ are made.
In Fig.\ref{fig:chi} we compare the time evolution of the exited
state population as predicted by both the post-Markovian master
equation [see Eqs. (\ref{eq:p1postMa}) and (\ref{eq:xi2})] and by
the memory kernel master equation [see Eqs. (\ref{eq:p1postMa})
and (\ref{eq:xi2memory})] with the exact dynamics [see Eq.
(\ref{eq:p1exact})], in correspondence to three different values
of the parameter $R=\gamma_0/\gamma$. We assume that the initial
state of the two-level system is the exited state.
From the figure one can see that while the memory kernel
approximated dynamics does violate positivity for strong enough
couplings ($R=5$ and $R=1$), the post-Markovian dynamics is always
positive. However, both the post-Markovian solution and the
second-order solution approximate well at all times $\tau$ the
exact dynamics only for small values of $R$, e.g. for small
couplings. Therefore, the specific example here considered shows
that there exist situations for which there is actually no
advantage in using the post-Markovian approach when compared to
second order perturbation theory. The reason why the
post-Markovian approach fails in describing the dynamics of the
system even for intermediate couplings is related to the way in
which such a master equation is derived. Let us recall the
physical meaning of the post-Markovian approach in terms of
generalized measurement interpretation. The derivation of the
post-Markovian master equation assumes that, after the time $t'$
at which the generalized measurement by the environment is
performed, the evolution of the system is Markovian. The time
distribution of the instants $t'$ at which the measurement is
performed is given by the memory kernel. Therefore one should
expect such master equation to be valid for $\tau_0 > \tau_R$.
Indeed, for $\tau_0 \leq \tau_R$ the assumption that the dynamics
after $t'$ is Markovian would not be well justified since the
reservoir correlation time would then be longer than the Markovian
dissipation time, and this would inevitably lead to a
non-negligible feedback of the environment to the system. We see
from Eq. (\ref{eq:p1exact})  that already for $R =\gamma_0/\gamma
\geq 0.5$ the exact dynamics of $P_1(\tau)$ shows an oscillatory
behavior. These oscillations may be seen, in a completely quantum
approach, as virtual absorption and re-emission of the same
quantum of energy from the environment. The description of these
quantum phenomena cannot be present in the post-Markovian
approach. It is exactly the appearance of these virtual exchanges
of energy which does not allow to give a physical interpretation
to the single trajectories for strongly non-Markovian systems,
since there seems to be no way for a single physical trajectory to
describe a virtual process.

\section{Summary and Conclusions}
In this paper we have taken into consideration two models of
generalized master equations with memory, and we have applied them
to the description of the non-Markovian dynamics of a qubit
interacting with a quantized bosonic reservoir in thermal
equilibrium. For the case of an exponential memory kernel we have
compared the solution of the recently proposed post-Markovian
master equation with the solution of the typical master equation
with memory kernel. We have demonstrated that, for the system
considered, the post-Markovian approach never violates positivity,
contrarily to the memory kernel master equation. We have then seen
that the memory kernel master equation coincides with the second
order expansion of the exact Nakajima-Zwanzig generalized master
equation. Since the memory kernel master equation is the limit of
the post-Markovian master equation for $\gamma_0 \ll \gamma$, it
is possible to give a generalized measurement interpretation to
the correlation function of the reservoir. Finally we have
considered the following physical implementation of the system:
the qubit describes the excited and ground electronic state of an
effective two-level atom crossing a high $Q$ cavity; the reservoir
is formed by the quantized modes of the high $Q$ cavity which are
distributed according to a Lorentzian peaked at the atomic Bohr
frequency. This physical system, typical of cavity QED, represents
one of the few examples of exactly solvable open quantum systems.
The comparison between the exact dynamics and the post-Markovian
and memory kernel solutions shows that there exist situations in
which the post-Markovian approach does not present any advantage
over the second order approximated memory kernel master equation.
The reason is traceable back to the fact that, by derivation, the
post-Markovian master equation cannot describe accurately
situations for which the characteristic time of the reservoir
$\tau_R$ is greater than the characteristic time of the system
$\tau_0$. When $\tau_R \ge \tau_0$ the dynamics is characterized
by virtual exchanges of energies between the system and the
environment which cannot be described by the post-Markovian
approach. Such virtual processes, absent in the Markovian
dynamics, seem to be responsible for the lack of a physical
interpretation of single quantum trajectories in terms of
continuous measurements performed by the environment.

\section{Acknowledgements}
S.M. thanks Nikolay Vitanov for the hospitality at the University
of Sofia, where part of the work was done, and acknowledges
financial support by the European Union's Transfer of Knowledge
project CAMEL (Grant No. MTKD-CT-2004-014427) and by the Angelo
Della Riccia Italian National Foundation.

\section*{Appendix A}
The damping basis method allows to solve the master equation given
by Eq.(\ref{eq:markovme}) by solving the eigenvalue equation
\begin{equation}
\mathcal{L} \rho_{\lambda} = \lambda \rho_{\lambda}.
\end{equation}
It turns out that the damping basis is \cite{Briegel93}
\begin{equation}
\rho_{\lambda_i}: \{\rho_{\lambda_1}=\sigma_0,
\rho_{\lambda_2}=\sigma_z, \rho_{\lambda_3} = \sigma_+,
\rho_{\lambda_4} = \sigma_- \}, \label{eq:dampingbasis}
\end{equation}
with $\sigma_0=\frac{1}{2}[I-\sigma_z/(2N+1)]$,
$\sigma_{\pm}=\sigma_x \pm i \sigma_y$, and $\sigma_x, \sigma_y,
\sigma_z$ the Pauli matrices. The corresponding eigenvalues are
\begin{equation}
\lambda_i: \{ \lambda_1 =0, \lambda_2= -2 \gamma_0 (N+1/2),
\lambda_3=\lambda_4= - \gamma_0 (N+1/2) \}. \label{eq:lambdai}
\end{equation}

The density matrix $\rho(t)$ can then be written, in the damping
basis, as follows
\begin{eqnarray}
\rho(t)= \sum_{\lambda_i} c_{\lambda_i} e^{\lambda_i t}
\rho{\lambda_i}, \label{eq:generalrho}
\end{eqnarray}
with $c_{\lambda_i} = {\rm Tr \left\{ \check{\rho}_{\lambda_i}
\rho(0) \right\}}$, where $\check{\rho}_{\lambda_i}$ is the dual
damping basis. Inserting the values of Eqs.
(\ref{eq:dampingbasis})-(\ref{eq:lambdai}) into Eq.
(\ref{eq:generalrho}), one gets
\begin{eqnarray}
\rho(t) &=& \frac{1}{2} \left\{ I - \left[ e^{-\gamma_0 (2N+1) t}
\left(
\rho_{11}-\rho_{22}+\frac{1}{2N+1}\right) \right. \right.\nonumber \\
&-& \left. \left.  \frac{1}{2N+1}\right] \sigma_z +
\frac{e^{-\gamma_0 (N+1/2) t}}{2} \left[ \rho_{12} \sigma_+ +
\rho_{21} \sigma_- \right] \right\}. \label{eq:solutionMarkovian}
\end{eqnarray}

\section*{Appendix B}
In this Appendix we recall the main steps to derive the general
solution of Eq. (\ref{eq:1}), as demonstrated in \cite{lidar05},
and we carry out the derivation for the case of a qubit
interacting with a quantized thermal reservoir. The initial step
to solve the post-Markovian master equation is the derivation of
the damping basis for the Markovian case (see Appendix A). As in
the previous Appendix, we denote with $\{\lambda_i\}$ the complex
eigenvalues and with $\{\rho_{\lambda_i}\}$ and
$\{\check{\rho}_{\lambda_i}\}$ the damping basis and its dual,
respectively. Then we write the density matrix as follows
\begin{equation}
\rho(t)= \mu_i(t) \rho_{\lambda_i}. \label{eq:3}
\end{equation}
Taking the Laplace transform of Eq.(\ref{eq:1}) one gets
\begin{equation}
s \tilde{\rho}(s)-\rho(0)= \left[\tilde{k}(s)\ast
\frac{\mathcal{L}}{s-\mathcal{L}} \right] \tilde{\rho}(s),
\label{eq:4}
\end{equation}
where $\ast$ denotes the convolution. Taking the Laplace transform
of Eq. (\ref{eq:3}) and using the previous equation one obtains
\begin{equation}
s \tilde{\mu}_i(s)-\mu_i(0)=\lambda_i \tilde{k}(s-\lambda_i)
\tilde{\mu}_i(s),
\end{equation}
and transforming back
\begin{equation}
\mu_i={\rm Lap}^{-1} \left[ \frac{1}{s-\lambda_i
\tilde{k}(s-\lambda_i)} \right] \mu_i(0) \equiv \xi_i(t) \mu_i(0).
\label{eq:mui}
\end{equation}
The coefficients $\xi_i$ may be calculated once fixed
$\mathcal{L}$ and $k(t)$ (see \cite{lidar05}), therefore one gets
\begin{equation}
\rho(t)=\sum_i \xi_i(t) \mu_i(0) \rho{\lambda_i}. \label{eq:rhomu}
\end{equation}

For the case of a qubit interacting with a $T$-temperature
reservoir, the damping basis is given by Eq.
(\ref{eq:dampingbasis}). Assuming an exponential memory kernel,
and using Eq. (\ref{eq:lambdai}), we have solved Eq.
(\ref{eq:mui}) obtaining
\begin{eqnarray}
\xi_1(t) &=& 1, \nonumber \\
\xi_2(t) &=& \xi(R,t), \nonumber \\
\xi_3(t) &=& \xi_4(t) = \xi(2R, t),
\end{eqnarray}
with $\xi(R,t)$ given by Eq. (\ref{eq:xi2}). Inserting the
previous equations into Eq. (\ref{eq:rhomu}), one gets Eq.
(\ref{eq:solutionpostM}).

\section*{Appendix C}
In this Appendix we demonstrate that the post-Markovian master
equation for a qubit never violates the positivity condition for
an exponential memory kernel. We have seen in Sec.
\ref{sec:positivity} that the positivity condition amounts at
$0\le \xi(R,\tau) \le 1$.

Let us firstly show that $\xi(R,\tau) \le 0$. By looking at Eq.
(\ref{eq:xipostM}), and remembering that for the zero temperature
case $|\lambda_2|=\gamma_0$, one sees immediately that this
corresponds to prove that
\begin{equation}
\left\{\begin{matrix}
  1-R > 0, & \hspace{1cm} e^{-R\tau} - R e^{-\tau} > 0; \\
  1-R < 0, & \hspace{1cm} e^{-R\tau} - R e^{-\tau} < 0.
\end{matrix}\right.
\end{equation}
The first set of inequalities is always satisfied since when $R<1$
then $e^{(1-R)\tau}>R$ at all times $\tau$. Similarly the second
set of inequalities is always satisfied since when $R>1$ then
$e^{-(R-1)\tau}\le 1 < R$ at all times $\tau$.

We now show that $\xi(R,\tau) \le 1$. Since $R>0$ and $\xi(R,\tau)
\ge 0 $ we have
\begin{eqnarray}
\xi(R,\tau) = \frac{e^{-R\tau} - R e^{-\tau}}{1-R} \le
\frac{e^{-\tau} - R e^{-\tau}}{1-R} = e^{- \tau} \le 1.
\end{eqnarray}


\begin{thebibliography}{99}
\bibitem{note1} A map $\Phi$ is completely positive iff satisfies
both $\Phi>0$ (positivity condition) and $\Phi \otimes I_n \ge 0
\forall n \in \mathbb{N}$, with $I_n$ the $n$-dimensional identity
operator.
\end{thebibliography}
\end{document}